\documentclass[aps,eqsecnum,12pt,nofootinbib]{revtex4}
\usepackage{amssymb}
\usepackage{bm}
\newcommand\comma{\ ,\ \ }

\begin{document}

\thispagestyle{empty}
\vspace*{-25mm}

\begin{flushright}
\baselineskip14pt
KAIST-TH/2005-09\\
YITP-05-26\\
KUNS-1973
\end{flushright}

\vskip8mm

\begin{center}
\Large\bf
A New $\bm{\delta N}$ Formalism for Multi-Component Inflation
\end{center}

\vspace{10mm}

\begin{center}
\large
Hyun-Chul Lee$^a$,
Misao Sasaki$^b$,
Ewan D. Stewart$^{a,c,d,}$\footnote{On sabbatical leave from Department of Physics,
KAIST, Daejeon, Republic of Korea},
Takahiro Tanaka$^e$\\
and Shuichiro Yokoyama$^e$
\end{center}

\vspace{5mm}

\begin{center}
{\em
$^a$Department of Physics, KAIST, Daejeon, Republic of Korea\\
$^b$YITP, Kyoto University, Kyoto, Japan\\
$^c$Department of Physics and Astronomy, University of Canterbury,\\
Christchurch, New Zealand\\
$^d$CITA, University of Toronto, Toronto, Canada\\
$^e$Department of Physics, Kyoto University, Kyoto, Japan
}
\end{center}

\begin{center}
(\today)
\end{center}

\vfill

\begin{center}
\Large Abstract
\end{center}

The $\delta N$ formula that relates the final curvature perturbation on comoving slices
to the inflaton perturbation on flat slices after horizon crossing is a powerful and
intuitive tool to compute the curvature perturbation spectrum from inflation.
However, it is customarily assumed further that the conventional slow-roll condition
is satisfied, and satisfied by all components, during horizon crossing.
In this paper, we develop a new $\delta N$ formalism for multi-component inflation
that can be applied in the most general situations.
This allows us to generalize the idea of general slow-roll inflation to
the multi-component case, in particular only applying the general slow-roll condition
to the relevant component.
We compute the power spectrum of the curvature perturbation in multi-component general slow-roll inflation, and find that under quite general conditions it is invertible.

\vfill
\newpage

\section{Introduction}

Thanks to recent advances in observational technologies,
an era of precision cosmology has commenced,
and there have been many studies attempting to determine,
or at least constrain, possible models/theories of the early universe
by cosmological observations. Notably, the analysis of
the first year WMAP data strongly indicates
that the universe is spatially flat, that the
primordial perturbation is adiabatic and Gaussian, and
that the spectrum is almost scale-invariant~\cite{WMAP}.
By and large, these support the now-standard picture
that our universe experienced an inflationary phase
at its early stage.
Furthermore, the PLANCK satellite, to be launched within
a couple of years from now, will provide us with
more detailed information about not only the CMB
temperature spectrum but also the polarization
spectrum~\cite{PLANCK}.

On the theoretical side, both supersymmetric particle theory and string theory
lead to the expectation that there are many scalar fields.
It is then important to understand the cosmological dynamics
of multi-component scalar fields and formulate a theory
of the cosmological perturbations generated from vacuum fluctuations
in those fields during inflation.

The inflationary dynamics of multi-component scalar fields
has been investigated by various authors from various aspects.
In particular, the $\delta N$ formalism \cite{SS,Starobinsky,ST},
in which the final curvature perturbation on comoving slices
is expressed as the $e$-folding number perturbation,
which in turn is given in terms of
the inflaton perturbation on flat slices after horizon crossing,
has proved to be a powerful tool to compute
the curvature perturbation spectrum from inflation.

However, most of the previous work \cite{SS,Nakamura,JO} has been based on the assumption
that all the fields satisfy the conventional slow-roll condition.
The conventional slow-roll condition assumes that the slow-roll parameters
are not only small but also slowly varying.
In the case of a single inflaton field, the slow-roll parameters should be small
in order to produce a scale-invariant spectrum but there can be cases in which
the slow-roll parameters are not slowly varying.
In such cases, the resulting power spectrum of the curvature perturbation
can be quite different from the conventional slow-roll spectrum.
Of course, since the slow-roll parameters are small, the spectrum is still almost
scale-invariant so that it is consistent with present observational data.
Thus it is important to relax the conventional slow-roll condition
and investigate the power spectrum of the curvature perturbation
generated from general slow-roll inflation \cite{Scott,gsr,JY,Minu},
which in the case of a single inflaton field means assuming the slow-roll parameters
are small but not necessarily slowly varying.

In the case of a multi-component inflaton, one has many slow-roll parameters but only the relevant ones need to be small to ensure an approximately scale-invariant spectrum,
\mbox{i.e.} one should only apply the general slow-roll condition
to the relevant component, with essentially no constraints on the others.
This makes the task of calculating a general formula for the spectrum challenging.

In this paper, we formulate the cosmological perturbation of a universe dominated by
a multi-component scalar field during inflation in a form as general as possible,
extend the $\delta N$ formalism developed previously \cite{SS}
by introducing a new quantity $\delta N$,
and derive a formula for the power spectrum of the curvature perturbation
in multi-component general slow-roll inflation.

This paper is organized as follows.
In Section~\ref{background},
we briefly review the $\delta N$ formalism \cite{SS},
on which our analysis relies heavily.
That is, we introduce the perturbation in the $e$-folding
number, $\delta\mathcal{N}$, and show that $\delta\mathcal{N}$
between two different perturbed spacelike hypersurfaces
is equal to the difference of the curvature perturbations
on the two hypersurfaces on super-horizon scales.
Taking the initial hypersurface to be flat and the final one to be comoving,
$\delta\mathcal{N}$ becomes equal to the curvature perturbation
on the final comoving hypersurface, $\mathcal{R}_\mathrm{c}$,
which is conserved in time on scales sufficiently greater than the horizon size
if the final hypersurface is taken at a sufficiently late time
that the perturbations have become adiabatic.

In Section~\ref{sms},
we first write down the perturbation equation
for a multi-component scalar field in the flat slicing.
We introduce a quantity $\delta N$ which closely resembles
the perturbation of the background $e$-folding number, $N$,
which we regard as a function of $\phi$ and $\dot\phi$.
We then derive an evolution equation for $\delta N$
valid from sub-horizon scales to super-horizon scales.
Finally, we show that this $\delta N$ can be identified with $\delta\mathcal{N}$
on super-horizon scales, hence with the curvature perturbation
on the comoving slice, $\mathcal{R}_\mathrm{c}$, at late times.

In Section~\ref{mfgsrps}, using the equations derived in the previous section,
we derive the general formula for the power spectrum of $\mathcal{R}_\mathrm{c}$
in multi-field general slow-roll inflation to leading order.
In Section~\ref{sac}, we conclude our paper.

Some mathematical formulae used for manipulations of the multi-component
perturbation equations and some connections with previous formulae
are given in the Appendices.

\section{$\bm{\delta\mathcal{N}}$}
\label{background}

Following Ref.~\cite{SS}, we introduce the perturbation
of the $e$-folding number $\delta\mathcal{N}$ and
show its relation to the curvature perturbation.

The background metric is
\begin{equation}
ds^2 = dt^2 - a(t)^2 \delta_{ij} \, dx^i dx^j ~.
\end{equation}
We define the $e$-folding number as
\begin{equation}
N \equiv \int_{t_\mathrm{fin}}^t H dt ~,
\end{equation}
where $H \equiv \dot{a}/a$
and $t_\mathrm{fin}$ is a late time when all trajectories have converged,
\mbox{i.e.} a time after complete reheating when the curvature perturbation
$\mathcal{R}_\mathrm{c}$ has become constant;
see Eq.~(\ref{deltaNformula}) below.

We write the scalar part of the perturbed metric as \cite{KS}
\begin{equation}\label{metricdef}
ds^2 = (1 + 2 A) dt^2 - 2 \partial_i B \, dt \, dx^i
- a^2 \left[ (1 + 2 \mathcal{R}) \delta_{ij}
 + 2 a^{-2} \partial_i \partial_j E \right] dx^i dx^j ~.
\end{equation}
The perturbed $e$-folding number is defined by
\begin{equation}\label{perturbedefold}
\mathcal{N} \equiv \int_{t_\mathrm{fin}}^{t} \frac{1}{3} \theta\, d\tau ~,
\end{equation}
where $\tau$ is the proper time, $d\tau = (1+A)dt$, and $\theta$ is the volume
expansion rate of the constant time hypersurfaces.
We write $\theta$ as
\begin{eqnarray}
\frac{1}{3} \theta \equiv H (1 + \mathcal{K}) ~.
\end{eqnarray}
In  the rest of this paper, we expand the perturbation variables
in terms of spatial scalar harmonics $Y_k(x^i)$ with the eigenvalue
$k^2$ \cite{KS}, \mbox{e.g.} $A(t,x^i)=\sum_k A_k(t)Y_k(x^i)$, and deal with the
harmonic coefficients by omitting the eigenvalue indices.
For example, we denote $A_k$ simply by $A$.

The perturbation of the expansion rate $\mathcal{K}$ is expressed
in terms of the metric perturbations in Eq.~(\ref{metricdef}) as
\begin{equation}
\mathcal{K} = - A + \frac{1}{H} \dot\mathcal{R} - \frac{q^2}{3H} \mathcal{S} ~,
\end{equation}
where $q^2 = k^2 / a^2$,
\begin{equation}
\mathcal{S} \equiv \dot{E} - 2 H E - B ~,
\end{equation}
and $\mathcal{R}$ describes the intrinsic
curvature perturbation of the constant time hypersurfaces,
and $\mathcal{S}$ the shear of the vector unit normal to them.

We note that, assuming the matter anisotropic stress is negligible,
$\mathcal{S}$ satisfies the equation,
\begin{eqnarray}
\dot{\mathcal{S}}+H\mathcal{S}=A+\mathcal{R} ~.
\label{sheareq}
\end{eqnarray}
Thus, apart from the modes of $\mathcal{S}$ sourced by $A+\mathcal{R}$
that remain regular in the limit $q^2\to0$,
$q^2\mathcal{S}$ decays rapidly as $a^{-3}$ once it is outside the horizon,
and this decaying behavior is independent of the choice of gauge.

The perturbation in the $e$-folding number is written as
\begin{equation}
\delta\mathcal{N}(t_\mathrm{fin},t) \equiv \mathcal{N} - N
= \mathcal{R}(t) -  \mathcal{R}(t_\mathrm{fin})
 - \frac{1}{3} \int_{t_\mathrm{fin}}^{t} q^2 \mathcal{S} \, dt ~.
\end{equation}
Taking the initial, $t = t_\mathrm{ini}$, hypersurface to be flat and the final,
$t = t_\mathrm{fin}$, hypersurface to be comoving
\begin{equation}
\delta\mathcal{N}(t_\mathrm{fin},t_\mathrm{ini})= - \mathcal{R}_\mathrm{c}(t_\mathrm{fin})
 - \frac{1}{3} \int_{t_\mathrm{fin}}^{t_\mathrm{ini}} q^2 \mathcal{S} \, dt ~.
\end{equation}
As discussed in the previous paragraph,
$q^2\mathcal{S}$ is either rapidly decaying or $\mathcal{S}$ is
regular in the limit $q^2 \rightarrow 0$. Hence,
the second term in the right hand side of this
 equation is negligible on scales sufficiently greater than the horizon.
So, taking the initial time $t_\mathrm{ini}$ sufficiently late that
this term is negligible, we have
\begin{equation}
\label{deltaNformula}
\delta\mathcal{N}(t_\mathrm{fin},t_\mathrm{ini}) \simeq
 - \mathcal{R}_\mathrm{c}(t_\mathrm{fin}) ~.
\end{equation}
This relation is valid irrespective of whether the background universe
 is dominated by scalar field or not.

\section{Multi-component scalar field during inflation}
\label{sms}

In the inflationary universe, the curvature perturbation
is generated from vacuum fluctuations in the scalar field
when the wavelengths of the Fourier modes are well inside the horizon.
Therefore generically it is necessary to solve the evolution
of these vacuum fluctuations until they are well outside the horizon.
However, for multi-component inflation, the curvature perturbation $\mathcal{R}_\mathrm{c}$
can still vary in time on super-horizon scales until the background trajectories
in the phase space converge to a single trajectory.
The convergence may occur only after the universe becomes totally radiation-dominated,
after which the background universe undergoes a universal evolution
and the only remaining modes of scalar-type perturbations are adiabatic ones
(we do not consider the case of isocurvature perturbations that persist until the present).
It is at this stage that $\mathcal{R}_\mathrm{c}$ becomes constant in time.

Thinking of the fact that what we need is only
the final value of the curvature perturbation, $\mathcal{R}_\mathrm{c}(t_\mathrm{fin})$,
it seems highly redundant to know the evolution of all the components.
In other words, if we can identify the part of the perturbations
that contributes to $\mathcal{R}_\mathrm{c}(t_\mathrm{fin})$,
we may solve only that part to obtain its value
without solving the full perturbation equations.

What the $\delta\mathcal{N}$ formula~(\ref{deltaNformula}) tells us
is that this final amplitude of the adiabatic perturbation
$\mathcal{R}_\mathrm{c}(t_\mathrm{fin})$ is given by
$\delta\mathcal{N}(t_\mathrm{ini},t_\mathrm{fin})$, \mbox{i.e.} the
perturbation in the $e$-folding number between the initial
flat hypersurface at $t=t_\mathrm{ini}$ during
inflation and the final comoving hypersurface at
$t=t_\mathrm{fin}$ when $\mathcal{R}_\mathrm{c}$ has relaxed
to a constant. Thus, the essential information we need is
$\delta\mathcal{N}$ on super-horizon scales during inflation.
Therefore, it will be convenient if we
can introduce a quantity that can be evolved from sub-horizon
scales to super-horizon scales and that can be identified
with $\delta\mathcal{N}$ there. The purpose of this section is
to introduce such a quantity which we denote by $\delta N$
in a multi-component scalar field dominated stage, and argue with several pieces
of strong, convincing evidence that it indeed matches to $\delta \mathcal{N}$
on super-horizon scales.
A rigorous proof of, including any necessary conditions for,
the equivalence between $\delta \mathcal{N}$ and $\delta N$
on super-horizon scales will be a topic of a subsequent paper~\cite{SSTY}.

The reason why we denote this quantity by $\delta N$ is that
its definition reduces to the perturbation of the background
$e$-folding number in the super-horizon limit.

\subsection{The background}

In this subsection, we review the evolution equations for the background,
and work out the properties of the $e$-folding number $N$ as a function
on the phase space of the scalar field. We denote the dimensions of the
field space by $D$.

\subsubsection{Basic background equations}

We assume that for $t \leq t_\mathrm{ini}$, \mbox{i.e.} while modes are leaving
 the horizon during inflation, we can take the action to be
\begin{equation}\label{action}
S = \int d^4x \sqrt{-g} \left[ - \frac{1}{2} R
 + \frac{1}{2} h_\mathbf{ab} g^{\mu\nu} \partial_\mu \phi^\mathbf{a} \partial_\nu
 \phi^\mathbf{b} - V(\phi) \right] ~.
\end{equation}
We do not require this effective description to continue to be valid for $t > t_\mathrm{ini}$.
Indeed, we expect that in most cases it will break down some time before $t_\mathrm{fin}$.

Here $h_{\mathbf{ab}}$ is the metric of the $D$-dimensional field space,
and following Wald~\cite{waldtext}, we adopt the abstract index notation,
\mbox{i.e.} we use boldface Latin indices $\mathbf{a},\mathbf{b},\ldots$
to denote tensors and their contractions in the field space,
while the first few Greek letters $\alpha,\beta,\ldots$ will be used
to label the components of tensors with respect to a basis of vectors $e_\alpha^\mathbf{a}$ and covectors $e^\alpha_\mathbf{a}$ in the field space, hence $\alpha=1,2,\ldots,D$.
Practically, one may regard the abstract index notation as the component notation
for an arbitrary basis.
One advantage of the abstract index notation is that it makes expressions involving derivatives in the curved field space concise.
Thus, for example, if we take $\phi^\alpha$ be the coordinates on the field space and use the usual mathematical definition of vectors as derivatives,
we have $v^\mathbf{a}\leftrightarrow v^\alpha\partial/\partial\phi^\alpha$
where $\partial/\partial\phi^\alpha$ is the coordinate basis.

The homogeneous background equations are
\begin{equation}\label{phieom}
\ddot\phi^\mathbf{a} + 3 H \dot\phi^\mathbf{a}
 + h^\mathbf{ab} \nabla_\mathbf{b} V(\phi) = 0 ~,
\end{equation}
\begin{equation}\label{Friedmann}
3 H^2 = \frac{1}{2} h_\mathbf{ab} \dot\phi^\mathbf{a} \dot\phi^\mathbf{b} + V ~,
\end{equation}
\begin{equation}
\dot{H} = - \frac{1}{2} h_\mathbf{ab} \dot\phi^\mathbf{a} \dot\phi^\mathbf{b} ~,
\end{equation}
where $\nabla_\mathbf{a}$ is the covariant derivative on the field space,
$\ddot\phi^\mathbf{a} \equiv D\dot\phi^\mathbf{a}/dt$,
and $D/dt$ is the covariant time derivative along a background
trajectory in the field space.

Because the background trajectories are determined by specifying both
$\phi^\alpha$ and $\dot\phi^\alpha$ in the field space, it is
convenient to extend the covariant derivatives to the phase space
in which $\phi^\alpha$ and $\dot\phi^\alpha$ are regarded as
independent coordinates.
The covariant derivatives in the phase space are defined in Appendix~\ref{cpd}.
We denote them by the subscripts $\phi^\mathbf{a}$ and $\dot\phi^\mathbf{a}$,
for example $Q_{\phi^\mathbf{a}}$ and $Q_{\dot\phi^\mathbf{a}}$,
and they are used extensively below.
However, readers who are not particularly interested
in the curved field space generalization may just regard them as
the standard partial derivatives with $\mathbf{a},\mathbf{b},\ldots$
being interpreted as the component indices.

\subsubsection{$N_{\phi^\mathbf{a}}$ and $N_{\dot\phi^\mathbf{a}}$}
\label{NphiNdotphi}

For $t \leq t_\mathrm{ini}$, we define that $N$ is represented in
phase space ($\phi,\dot\phi$) as
\begin{equation}
N(\phi,\dot\phi) \equiv \int_{t_\mathrm{fin}}^{t(\phi,\dot\phi)} H dt ~,
\end{equation}
where the integral is performed along the trajectory that passes through $(\phi,\dot\phi)$.
Therefore, we can write the Hubble parameter as
\begin{equation}\label{Ndot}
N_{\phi^\mathbf{b}} \dot\phi^\mathbf{b}
 + N_{\dot\phi^\mathbf{b}} \ddot\phi^\mathbf{b} = H ~,
\end{equation}
where the covariant partial derivatives are defined in Appendix~\ref{cpd}.
Differentiating this equation with respect to $\phi^{\mathbf{a}}$
 and $\dot\phi^{\mathbf{a}}$ respectively, we obtain
\begin{eqnarray}
\frac{D N_{\phi^\mathbf{a}}}{dt}
& = & - N_{\dot\phi^\mathbf{b}} \left(\ddot\phi^\mathbf{b}\right)_{\phi^\mathbf{a}}
 - N_{\dot\phi^\mathbf{b}} {R^\mathbf{b}}_\mathbf{cda} \dot\phi^\mathbf{c}
 \dot\phi^\mathbf{d} + H_{\phi^\mathbf{a}}
\nonumber \\ \label{evolutionN1}
& = & N_{\dot\phi^\mathbf{b}} h^\mathbf{bc} V_{\phi^\mathbf{c}\phi^\mathbf{a}}
 + 3 N_{\dot\phi^\mathbf{b}} \dot\phi^\mathbf{b} H_{\phi^\mathbf{a}}
 - N_{\dot\phi^\mathbf{b}} {R^\mathbf{b}}_\mathbf{cda} \dot\phi^\mathbf{c}
 \dot\phi^\mathbf{d} + H_{\phi^\mathbf{a}} ~,
\\
\frac{D N_{\dot\phi^\mathbf{a}}}{dt}
& = & - N_{\phi^\mathbf{a}} - N_{\dot\phi^\mathbf{b}}
 \left(\ddot\phi^\mathbf{b}\right)_{\dot\phi^\mathbf{a}}+ H_{\dot\phi^\mathbf{a}}
\nonumber \\ \label{evolutionN2}
& = & - N_{\phi^\mathbf{a}} + 3 H N_{\dot\phi^\mathbf{a}}
 + 3 N_{\dot\phi^\mathbf{b}} \dot\phi^\mathbf{b} H_{\dot\phi^\mathbf{a}}
 + H_{\dot\phi^\mathbf{a}} ~,
\end{eqnarray}
where ${R^\mathbf{b}}_\mathbf{cda}$ is the scalar field space curvature tensor. From Eq.~(\ref{Friedmann}), we have
\begin{equation}
H_{\phi^\mathbf{a}} = \frac{V_{\phi^\mathbf{a}}}{6 H}
\comma
H_{\dot\phi^\mathbf{a}} = \frac{h_\mathbf{ab} \dot\phi^\mathbf{b}}{6 H} ~.
\end{equation}
Using these equations, we obtain the evolution equations for
 $N_{\dot\phi^\mathbf{b}}$ and $N_{\phi^\mathbf{b}}$ as
\begin{eqnarray}
\lefteqn{
\frac{D^2}{dt^2}\left( \frac{h^\mathbf{ab} N_{\dot\phi^\mathbf{b}}}{a^3} \right)
 + 3 H \frac{D}{dt}\left( \frac{h^\mathbf{ab} N_{\dot\phi^\mathbf{b}}}{a^3} \right)
 - R^\mathbf{a}_\mathbf{\ bcd} \dot\phi^\mathbf{b} \dot\phi^\mathbf{c}
 \left(\frac{h^\mathbf{de} N_{\dot\phi^\mathbf{e}}}{a^3}\right)
 + h^\mathbf{ab} V_{\phi^\mathbf{b}\phi^\mathbf{c}}
 \left(\frac{h^\mathbf{cd} N_{\dot\phi^\mathbf{d}}}{a^3}\right)
} \nonumber \\ \label{Ndotphieom}
& = & \frac{1}{a^3} \frac{D}{dt}
\left( \frac{a^3 \dot\phi^\mathbf{a} \dot\phi^\mathbf{b}}{H} \right)
 h_\mathbf{bc} \left( \frac{h^\mathbf{cd} N_{\dot\phi^\mathbf{d}}}{a^3} \right)
 + \frac{1}{3 a^3} \frac{D}{dt}\left( \frac{\dot\phi^\mathbf{a}}{H} \right)
\hspace{35ex}
\end{eqnarray}
and
\begin{equation}\label{Nphieom}
\frac{h^\mathbf{ab} N_{\phi^\mathbf{b}}}{a^3}
 = - \frac{D}{dt}\left( \frac{h^\mathbf{ab} N_{\dot\phi^\mathbf{b}}}{a^3} \right)
 + \frac{N_{\dot\phi^\mathbf{b}} \dot\phi^\mathbf{b} \dot\phi^\mathbf{a}}{2Ha^3}
 + \frac{\dot\phi^\mathbf{a}}{6 H a^3} ~.
\end{equation}
Here, Eq.~(\ref{Ndotphieom}) has the particular solution
\begin{equation}\label{particular}
\frac{h^\mathbf{ab} N_{\dot\phi^\mathbf{b}}}{a^3}
 = \frac{\dot\phi^\mathbf{a}}{6 H} \int \frac{dt}{a^3} ~.
\end{equation}
The following contraction, derived from Eq.~(\ref{evolutionN2}) using Eq.~(\ref{Ndot}), will also be useful
\begin{equation}
\label{Ndotphidotphi}
3 N_{\dot\phi^\mathbf{a}} \dot\phi^\mathbf{a} = 1 + \frac{1}{3} \frac{\dot{H}}{H^2}
+ \frac{D N_{\dot\phi^\mathbf{a}}}{dt} \frac{\dot\phi^\mathbf{a}}{H}
- N_{\dot\phi^\mathbf{a}} \frac{D}{dt} \left(\frac{\dot\phi^\mathbf{a}}{H}\right) ~.
\end{equation}

\subsection{Perturbations}

In this subsection, we review the perturbation equations for a universe dominated
by a multi-component scalar field, and introduce a quantity $\delta N$ which we will
identify with $\delta\mathcal{N}$ on super-horizon scales and whose sub-horizon to
super-horizon evolution is particularly simple.

\subsubsection{$\delta \phi$ on flat slices}

During inflation, the metric perturbations $A$ and $\mathcal{S}$ satisfy
\begin{eqnarray}\label{A}
A & = & \frac{1}{2} h_\mathbf{ab}
 \frac{\dot\phi^\mathbf{a}}{H} \delta\phi^\mathbf{b} + \frac{\dot\mathcal{R}}{H} ~,
\\
q^2 \mathcal{S} & = & \frac{1}{2} h_\mathbf{ab} \left[ \frac{D}{dt}
 \left( \frac{\dot\phi^\mathbf{a}}{H} \right) \delta\phi^\mathbf{b}
 - \frac{\dot\phi^\mathbf{a}}{H} \dot{\delta\phi^\mathbf{b}} \right]
 - \frac{\dot{H}}{H^2} \dot\mathcal{R} + \frac{q^2}{H} \mathcal{R} ~.
\end{eqnarray}

We will use a subscript $\mathrm{f}$ to denote a quantity evaluated on flat hypersurfaces,
\mbox{i.e.} in the gauge $\mathcal{R}=0$.
In particular,
\begin{eqnarray}
\delta\phi_\mathrm{f}^\mathbf{a}
 &=& \delta\phi^\mathbf{a} - \frac{\dot\phi^\mathbf{a}}{H} \mathcal{R} ~,
\\
\mathcal{S}_\mathrm{f} &=& \mathcal{S} - \frac{1}{H} \mathcal{R} ~.
\end{eqnarray}
These variables are gauge invariant.

The scalar field perturbations on flat hypersurfaces satisfy~\cite{SS}
\begin{equation}\label{deltaphieom}
\ddot{\delta\phi_\mathrm{f}^\mathbf{a}} + 3H \dot{\delta\phi_\mathrm{f}^\mathbf{a}}
 - R^\mathbf{a}_\mathbf{\ bcd} \dot\phi^\mathbf{b} \dot\phi^\mathbf{c}
 \delta\phi^\mathbf{d}_\mathrm{f} + q^2 \delta\phi_\mathrm{f}^\mathbf{a}
+ h^\mathbf{ab} V_{\phi^\mathbf{b}\phi^\mathbf{c}} \delta\phi_\mathrm{f}^\mathbf{c}
= \frac{1}{a^3} \frac{D}{dt}
\left( \frac{a^3 \dot\phi^\mathbf{a} \dot\phi^\mathbf{b}}{H} \right)
 h_\mathbf{bc} \delta\phi_\mathrm{f}^\mathbf{c} ~.
\end{equation}
In the limit $q^2 \rightarrow 0$, Eq.~(\ref{deltaphieom}) has the solution
\begin{equation}\label{SHGM}
\delta\phi_\mathrm{f}^\mathbf{a} \propto \frac{\dot\phi^\mathbf{a}}{H} ~,
\label{growingmode}
\end{equation}
corresponding to the super-horizon adiabatic growing mode.
Comparing with Eqs.~(\ref{Ndotphieom}) and~(\ref{particular}),
 we see that, in the limit $q^2 \rightarrow 0$,
 Eq.~(\ref{deltaphieom}) also has the solution
\begin{equation}\label{SHDM1}
\delta\phi_\mathrm{f}^\mathbf{a}
 \propto \frac{h^\mathbf{ab} N_{\dot\phi^\mathbf{b}}}{a^3}
 - \frac{\dot\phi^\mathbf{a}}{6 H} \int^t_{t_*}\frac{dt}{a^3} ~,
\end{equation}
and, using Eq.~(\ref{Nphieom}),
\begin{equation}\label{SHDM2}
\dot{\delta\phi_\mathrm{f}^\mathbf{a}} \propto
 - \frac{h^\mathbf{ab} N_{\phi^\mathbf{b}}}{a^3}
 + \frac{N_{\dot\phi^\mathbf{b}} \dot\phi^\mathbf{b} \dot\phi^\mathbf{a}}{2Ha^3}
- \frac{1}{6} \frac{D}{dt}\left( \frac{\dot\phi^\mathbf{a}}{H} \right)
 \int^t_{t_*}\frac{dt}{a^3} ~.
\end{equation}
Taking the lower bound of the time integral, $t_*$, in the above equations
to be a sufficiently late time, this solution describes a super-horizon decaying mode.
Note that a different choice of $t_*$ corresponds to adding the
growing mode solution~(\ref{growingmode}) to it.
Note also that the time integral is defined even after inflation when the universe
is no longer dominated by the scalar field.
Thus as a simplest choice we set $t_*=t_\mathrm{fin}$,
the time at which we evaluate the final amplitude of the
 curvature perturbation $\mathcal{R}_\mathrm{c}$.

\subsubsection{Wronskian}
\label{Wronskian}

The Wronskian, $W$, of solutions $\delta\phi_{\mathrm{f},1}$ and
$\delta\phi_{\mathrm{f},2}$ of Eq.~(\ref{deltaphieom}) is defined as
\begin{equation}
W(\delta\phi_{\mathrm{f},2},\delta\phi_{\mathrm{f},1})
 \equiv h_\mathbf{ab} \left( \dot{\delta\phi^\mathbf{a}_{\mathrm{f},2}}
 \, \delta\phi^\mathbf{b}_{\mathrm{f},1}
 - \delta\phi^\mathbf{a}_{\mathrm{f},2} \,
 \dot{\delta\phi^\mathbf{b}_{\mathrm{f},1}} \right) ~,
\end{equation}
and has the property
\begin{equation}
W  \propto \frac{1}{a^3} ~.
\end{equation}

We also define the super-horizon Wronskian,
$W_0(\delta\phi_{\mathrm{f},0},\delta\phi_{\mathrm{f}})$,
as the Wronskian of a solution $\delta\phi_{\mathrm{f}}$ of Eq.~(\ref{deltaphieom})
and a super-horizon solution $\delta\phi_{\mathrm{f},0}$ of Eq.~(\ref{deltaphieom}),
\mbox{i.e.} a solution of Eq.~(\ref{deltaphieom}) in the limit $q^2 \rightarrow 0$.
Then
\begin{equation}\label{dotW0}
\frac{D}{dt} \left( a^3 W_0 \right)
 = q^2 \left( a^3 h_\mathbf{ab} \delta\phi^\mathbf{a}_{\mathrm{f},0} \right)
 \delta\phi^\mathbf{b}_{\mathrm{f}}
\end{equation}
and
\begin{equation}\label{ddotW0}
\frac{D^2}{dt^2} \left( a^3 W_0 \right)
+ 5 H \frac{D}{dt} \left( a^3 W_0 \right) + q^2 \left( a^3 W_0 \right)
 = 2 q^2 \frac{D}{dt} \left( a^3 h_\mathbf{ab}
 \delta\phi^\mathbf{a}_{\mathrm{f},0} \right) \delta\phi^\mathbf{b}_{\mathrm{f}} ~.
\end{equation}
These evolution equations for $a^3W_0$ motivate our final choice
for $\delta N_\mathrm{f}$, Eq.~(\ref{deltaNXflat}), introduced in the next subsection.

Taking the super-horizon decaying mode of Eqs.~(\ref{SHDM1}) and~(\ref{SHDM2}),
\begin{equation}
\delta\phi_\mathrm{f,d}^\mathbf{a}
 = - \frac{h^\mathbf{ab} N_{\dot\phi^\mathbf{b}}}{a^3}
 + \frac{\dot\phi^\mathbf{a}}{6 H} \int^t_{t_\mathrm{fin}} \frac{dt}{a^3} ~,
\end{equation}
and
\begin{eqnarray}
\dot{\delta\phi_\mathrm{f,d}^\mathbf{a}}
 & = & \frac{h^\mathbf{ab} N_{\phi^\mathbf{b}}}{a^3}
 - \frac{N_{\dot\phi^\mathbf{b}} \dot\phi^\mathbf{b} \dot\phi^\mathbf{a}}{2Ha^3}
+ \frac{1}{6} \frac{D}{dt}
\left( \frac{\dot\phi^\mathbf{a}}{H} \right) \int^t_{t_\mathrm{fin}} \frac{dt}{a^3} ~,
\end{eqnarray}
we get the super-horizon Wronskian
\begin{eqnarray}
\lefteqn{
a^3 W_0(\delta\phi_\mathrm{f,d},\delta\phi_\mathrm{f}) }
\nonumber\\
& = & N_{\phi^\mathbf{a}} \delta\phi_\mathrm{f}^\mathbf{a}
 + N_{\dot\phi^\mathbf{a}} \dot{\delta\phi_\mathrm{f}^\mathbf{a}}
 - \frac{1}{2H} N_{\dot\phi^\mathbf{a}}
 \dot\phi^\mathbf{a} \dot\phi_\mathbf{b} \delta\phi_\mathrm{f}^\mathbf{b}
+ \frac{1}{6} h_\mathbf{ab}
 \left[ \frac{D}{dt} \left( \frac{\dot\phi^\mathbf{a}}{H} \right)
 \delta\phi_\mathrm{f}^\mathbf{b}
 - \frac{\dot\phi^\mathbf{a}}{H} \dot{\delta\phi_\mathrm{f}^\mathbf{b}} \right]
 a^3 \int^t_{t_\mathrm{fin}} \frac{dt}{a^3}
\nonumber \\ \label{Wzero}
& = & N_{\phi^\mathbf{a}} \delta\phi_\mathrm{f}^\mathbf{a}
 + N_{\dot\phi^\mathbf{a}} \left( \dot{\delta\phi_\mathrm{f}^\mathbf{a}}
 - \dot\phi^\mathbf{a} A_\mathrm{f} \right)
 + \frac{1}{3} q^2 \left( a^3 \int^t_{t_\mathrm{fin}} \frac{dt}{a^3} \right)
 \mathcal{S}_\mathrm{f} ~.
\end{eqnarray}

\subsubsection{$\delta N$}
\label{secdeltaN}

In this section we introduce a quantity $\delta N$ which has the key properties
that it behaves as $\delta\mathcal{N}$ on super-horizon scales,
but is locally defined apart from the background function $N$
and so is straightforward to evaluate.\footnote{Strictly speaking,
 our final choice for $X$ in Eq.~(\ref{Xflat}) is not local,
but it only depends on background quantities and is
 well approximated by a local quantity.}
Although not mandatory, we also look for $\delta N$
whose sub-horizon to super-horizon evolution becomes as simple
as possible.

With this motivation, we define
\footnote{See Appendix~\ref{AppConfig} for the relationship between this definition and the configuration space formula used in Ref.~\cite{SS}.}
\begin{eqnarray}\label{deltaNbb}
\delta N & \equiv & \mathbb{N}_{\phi^\mathbf{a}} \delta\phi^\mathbf{a}
 + \mathbb{N}_{\dot\phi^\mathbf{a}} \dot{\delta\phi^\mathbf{a}}
\\ \label{deltaN}
& \equiv & N_{\phi^\mathbf{a}} \delta\phi^\mathbf{a}
 + N_{\dot\phi^\mathbf{a}}
 \left( \dot{\delta\phi^\mathbf{a}} - \dot\phi^\mathbf{a} A \right)
 - \frac{q^2}{9H} X \mathcal{S}_\mathrm{f} ~,
\end{eqnarray}
where
\begin{equation}
\mathbb{N}_{\phi^\mathbf{a}} = N_{\phi^\mathbf{a}}
- \frac{1}{2H} N_{\dot\phi^\mathbf{b}} \dot\phi^\mathbf{b} \dot\phi_\mathbf{a}
 - \frac{X}{18 H} \frac{D}{dt} \left( \frac{\dot\phi_\mathbf{a}}{H} \right)
\end{equation}
and
\begin{equation}
\mathbb{N}_{\dot\phi^\mathbf{a}} = N_{\dot\phi^\mathbf{a}}
 + \frac{X}{18 H} \frac{\dot\phi_\mathbf{a}}{H} ~,
\end{equation}
with $X$ being some, as yet unspecified, function of $t$.

There are several possibilities for the choice of $X$.
The choice $X = 0$ gives
\begin{equation}\label{deltaNX0}
\delta N = N_{\phi^\mathbf{a}} \delta\phi^\mathbf{a}
+ N_{\dot\phi^\mathbf{a}} \left( \dot{\delta\phi^\mathbf{a}}
 - \dot\phi^\mathbf{a} A \right)
\end{equation}
and
\begin{equation}
\mathbb{N}_{\dot\phi^\mathbf{a}} = N_{\dot\phi^\mathbf{a}} ~.
\end{equation}
Note that we may interpret
\begin{equation}
\dot{\delta\phi^\mathbf{a}} - \dot\phi^\mathbf{a} A
 = \delta \left( \frac{d\phi^\mathbf{a}}{d\tau} \right) ~.
\label{deltadotphi}
\end{equation}
Hence $\delta N$ may be regarded as the perturbation of the background $e$-folding number,
provided that $\left\{\delta\phi^\mathbf{a},\delta(d\phi^\mathbf{a}/d\tau)\right\}$
is regarded as the perturbation of the background field, \mbox{i.e.} the difference
between two adjacent background trajectories in the phase space.
This choice corresponds to the simplest extrapolation from the background $\delta N$
with $(\phi,\dot\phi)$ as the phase space variables.
Note also that, from Eqs.~(\ref{dotphitopi3}) and~(\ref{deltaPi}),
one has a similar interpretation if one chooses $(\phi,\pi)$ as the phase space variables,
where $\pi^\mathbf{a} \equiv \dot\phi^\mathbf{a}/H$,
provided that $\left\{\delta\phi^\mathbf{a},\delta(d\phi^\mathbf{a}/d\mathcal{N})\right\}$
is regarded as the perturbation of the background field.

Another choice may be motivated by taking $(\phi,\pi)$ as the phase space variables
with $\delta\pi^\mathbf{a} \equiv \dot{\delta\phi^\mathbf{a}}/H$.
Eq.~(\ref{dotphitopi2}) then suggests the choice
$X = 3 N_{\pi^\mathbf{a}} \pi^\mathbf{a}$.
This gives
\begin{equation}\label{deltaNXpi}
\delta N = N_{\phi^\mathbf{a}} \delta\phi^\mathbf{a}
 + N_{\pi^\mathbf{a}} \left( \delta\pi^\mathbf{a}
 - \pi^\mathbf{a} \frac{\dot\mathcal{R}}{H} \right)
\end{equation}
and
\begin{equation}
\mathbb{N}_{\dot\phi^\mathbf{a}} = \frac{1}{H} N_{\pi^\mathbf{a}} ~.
\end{equation}

Our final choice is
\begin{equation}\label{Xflat}
X = - 3 H a^3 \int_{t_\mathrm{fin}}^t \frac{dt}{a^3}
 = 1 + \mathcal{O} \left( \frac{\dot{H}}{H^2} \right) ~,
\end{equation}
relevant in the case of flat slicing,
and motivated by Eq.~(\ref{Wzero}) for the super-horizon Wronskian $a^3W_0$.
With this choice of $X$, we have
\begin{equation}\label{deltaNXflat}
\delta N = N_{\phi^\mathbf{a}} \delta\phi^\mathbf{a}
 + N_{\dot\phi^\mathbf{a}} \left( \dot{\delta\phi^\mathbf{a}}
 - \dot\phi^\mathbf{a} A \right)
 + \frac{1}{3} q^2 \left( a^3 \int^t_{t_\mathrm{fin}} \frac{dt}{a^3} \right)
 \mathcal{S}_\mathrm{f} ~,
\end{equation}
\begin{equation}
\mathbb{N}_{\dot\phi^\mathbf{a}} = N_{\dot\phi^\mathbf{a}}
 - \frac{1}{6} \left( a^3 \int^t_{t_\mathrm{fin}} \frac{dt}{a^3} \right) h_\mathbf{ab}
 \frac{\dot\phi^\mathbf{b}}{H}
\end{equation}
and
\begin{equation}
\frac{\mathbb{N}_{\phi^\mathbf{a}}}{a^3}
 = - \frac{D}{dt} \left( \frac{\mathbb{N}_{\dot\phi^\mathbf{a}}}{a^3} \right) ~.
\end{equation}

The choice of $X$ is irrelevant on super-horizon scales,
as long as $X$ does not grow rapidly in time,
but it affects the evolution of $\delta N$ from sub-horizon to super-horizon scales.
We will use this freedom to optimize this evolution, but for the moment
let us leave the function $X$ in $\delta N$ unspecified.

Choosing flat hypersurfaces, we have
\begin{equation}
{\delta N}_\mathrm{f} = \delta N - \mathcal{R} ~.
\end{equation}
Using the gauge transformation property of $\delta N$,
one can readily show that this is a gauge-invariant quantity.
Taking the time derivative of $\delta N_\mathrm{f}$, we find
\begin{equation}
\dot{\delta N}_\mathrm{f}
 = - q^2 \mathbb{N}_{\dot\phi^\mathbf{a}} \delta\phi_\mathrm{f}^\mathbf{a}
 - \frac{1}{3} q^2 \left[ 1 - \left( 1 - \frac{\epsilon}{3} \right) X
 + \frac{1}{3H} \dot{X} \right] \mathcal{S}_\mathrm{f} ~,
\label{deltaNdot}
\end{equation}
where $\epsilon \equiv -\dot{H}/H^2$ is one of the slow-roll parameters.
Thus ${\delta N}_\mathrm{f}$ is constant on super-horizon scales.
In the special case that our scalar field description is still valid
 at $t=t_\mathrm{fin}$, \mbox{i.e.} if it is valid beyond the point where
 convergence of trajectories occurs, we can evaluate the constant by
 inserting the super-horizon adiabatic growing mode, Eq.~(\ref{SHGM}),
 into the definition of $\delta N$, Eq.~(\ref{deltaN}), to get
\begin{equation}
{\delta N}_\mathrm{f}(t_\mathrm{ini}) = {\delta N}_\mathrm{f}(t_\mathrm{fin})
= \left. \frac{H \dot\phi_\mathbf{a} \delta\phi_\mathrm{f}^\mathbf{a}}
    {\dot\phi_\mathbf{b} \dot\phi^\mathbf{b}} \right|_{t=t_\mathrm{fin}}
= - \mathcal{R}_\mathrm{c}(t_\mathrm{fin}) ~.
\end{equation}
Comparing this with $\delta\mathcal{N}$ given in Eq.~(\ref{deltaNformula}),
we see that
$\delta N_\mathrm{f}(t_\mathrm{ini}) = \delta\mathcal{N}(t_\mathrm{fin},t_\mathrm{ini})$,
where $t_\mathrm{ini}$ is a time when the mode is sufficiently far outside the horizon.
Thus we have established the equivalence of $\delta N_\mathrm{f}$ and
$\delta\mathcal{N}$ on super-horizon scales in this special case,
and the natural form of Eqs.~(\ref{deltaNX0}) and~(\ref{deltaNXpi}) and their equivalence convinces us that this is true generally.
We aim to prove this rigorously in a subsequent paper~\cite{SSTY}.

Taking the derivative again we get the equation of motion for ${\delta N}_\mathrm{f}$
\begin{eqnarray}
\lefteqn{
\ddot{\delta N}_\mathrm{f} + 5 H \, \dot{\delta N}_\mathrm{f}+ q^2 {\delta N}_\mathrm{f}
 = - 2 q^2 \dot\mathbb{N}_{\dot\phi^\mathbf{a}} \delta\phi_\mathrm{f}^\mathbf{a}
} \nonumber \\ \label{eomdeltaN} && \mbox{}
- \frac{2}{3} H q^2 \left[ 1 - \left( 1 - \frac{\epsilon}{3}
 - \frac{\dot\epsilon}{6H} \right) X - \frac{1-2\epsilon}{6H} \dot{X}
 + \frac{1}{6H^2} \ddot{X} \right] \mathcal{S}_\mathrm{f} ~.
\end{eqnarray}

Now let us fix the function $X$.
As we may expect from the evolution equations for the super-horizon Wronskian $a^3W_0$,
Eqs.~(\ref{dotW0}) and (\ref{ddotW0}),
if we choose $X$ as in Eq.~(\ref{Xflat}), the $\mathcal{S}_\mathrm{f}$ terms
in Eqs.~(\ref{deltaNdot}) and (\ref{eomdeltaN}) disappear:
\begin{equation}\label{deltaNFdot}
\dot{\delta N}_\mathrm{f}
 = - q^2 \mathbb{N}_{\dot\phi^\mathbf{a}} \delta\phi_\mathrm{f}^\mathbf{a}
\end{equation}
and
\begin{equation}\label{deltaNFddot}
\ddot{\delta N}_\mathrm{f} + 5 H \, \dot{\delta N}_\mathrm{f}
+ q^2 {\delta N}_\mathrm{f}
= - 2 q^2 \dot\mathbb{N}_{\dot\phi^\mathbf{a}} \delta\phi_\mathrm{f}^\mathbf{a} ~.
\end{equation}

What we want to do is to solve the above equation
from well inside the horizon to well outside the horizon,
in particular to the time $t=t_\mathrm{ini}$ at which $\delta N_\mathrm{f}$
and $\delta\mathcal{N}$ can be identified with each other.
The left hand side of Eq.~(\ref{deltaNFddot}) gives a simple evolution equation
for the single quantity we are interested in, but the right hand side is not so simple.
In general, we would need to know the solution for $\delta\phi_\mathrm{f}^\mathbf{a}$.
However, then we would not need to solve Eq.~(\ref{deltaNFddot}) in the first place,
because $\delta N_\mathrm{f}$ can be readily constructed from
$\delta\phi_\mathrm{f}^\mathbf{a}$.
Therefore, we look for situations in which the whole knowledge of
$\delta\phi_\mathrm{f}^\mathbf{a}$ may not be needed.
For this purpose, we decompose $\delta\phi_\mathrm{f}^\mathbf{a}$ into relevant and irrelevant components,
\begin{equation}
\delta\phi_\mathrm{f}^\mathbf{a}
 = \delta\phi_\parallel^\mathbf{a} + \delta\phi_\bot^\mathbf{a} ~,
\end{equation}
with the relevant direction defined by $\mathbb{N}_{\dot\phi^\mathbf{a}}$,
\begin{equation}\label{PQdeltaphi}
\delta\phi_\parallel^\mathbf{a} \equiv P^\mathbf{a}_\mathbf{b} \,
 \delta\phi_\mathrm{f}^\mathbf{b}
\comma
\delta\phi_\bot^\mathbf{a} \equiv Q^\mathbf{a}_\mathbf{b} \,
 \delta\phi_\mathrm{f}^\mathbf{b}
\end{equation}
where
\begin{equation}
P^\mathbf{a}_\mathbf{b} \equiv
 \frac{ h^\mathbf{ac} \mathbb{N}_{\dot\phi^\mathbf{c}}
 \mathbb{N}_{\dot\phi^\mathbf{b}}}
{h^\mathbf{de} \mathbb{N}_{\dot\phi^\mathbf{d}} \mathbb{N}_{\dot\phi^\mathbf{e}}}
\comma
Q^\mathbf{a}_\mathbf{b} \equiv \delta^\mathbf{a}_\mathbf{b} - P^\mathbf{a}_\mathbf{b} ~,
\end{equation}
so that, from Eq.~(\ref{deltaNFdot}), $\delta\phi_\parallel^\mathbf{a}$ can be expressed entirely in terms of ${\delta N}_\mathrm{f}$.
Then Eq.~(\ref{deltaNFddot}) can be rewritten as
\begin{equation}\label{deltaNeom}
\ddot{\delta N}_\mathrm{f} + 5 H \, \dot{\delta N}_\mathrm{f}
+ q^2 {\delta N}_\mathrm{f}
= 2 \left( \frac{h^\mathbf{ab} \dot\mathbb{N}_{\dot\phi^\mathbf{a}} \mathbb{N}_{\dot\phi^\mathbf{b}}} {h^\mathbf{cd} \mathbb{N}_{\dot\phi^\mathbf{c}} \mathbb{N}_{\dot\phi^\mathbf{d}}} \right) \dot{\delta N}_\mathrm{f} - 2 q^2 \dot\mathbb{N}_{\dot\phi^\mathbf{a}} \delta\phi_\bot^\mathbf{a} ~.
\end{equation}
Thus, in situations when the contribution of the $\delta\phi_\bot^\mathbf{a}$ term
can be neglected, the equation forms a closed system,
making it possible to solve for $\delta N_\mathrm{f}$
without the full knowledge of the scalar field perturbations.
Furthermore, we can use observational constraints on the spectrum to force
${\delta N}_\mathrm{f}$ to obey the general slow-roll condition,
while we have no real constraints on $\delta\phi_\bot^\mathbf{a}$
apart from that its contribution should be small.

Before closing this section, let us further rewrite Eq.~(\ref{deltaNeom}) in a form
suitable for explicit calculations.
Changing time variable to $x = k \xi$, where $\xi = - \int dt/a$ is minus
the conformal time, we obtain
\begin{equation}\label{deltaNeomx}
{\delta N}_\mathrm{f}'' - \frac{4}{x} \, {\delta N}_\mathrm{f}' + {\delta N}_\mathrm{f}
= \frac{2}{x} \frac{\Pi'}{\Pi} \, {\delta N}_\mathrm{f}'
- 2 \dot\mathbb{N}_{\dot\phi^\mathbf{a}} \delta\phi_\bot^\mathbf{a}
\end{equation}
where
\begin{equation}\label{Pi2}
\Pi^2 \equiv \left( \frac{3}{2\pi a^2 \xi^2} \right)^2 h^\mathbf{ab} \mathbb{N}_{\dot\phi^\mathbf{a}} \mathbb{N}_{\dot\phi^\mathbf{b}} ~,
\end{equation}
${\delta N}_\mathrm{f}' \equiv d\,{\delta N}_\mathrm{f} / dx$,
$\Pi' \equiv d\Pi / d\ln \xi$,
and $a\xi \simeq 1/H$ for $\epsilon \ll 1$.
This form of the equation for ${\delta N}_\mathrm{f}$ will be used in the next section.

\section{Multi-field general slow-roll power spectrum}
\label{mfgsrps}

At sufficiently early times when all cosmologically relevant wavelengths
are well inside the horizon scale, we assume that the time scale for the
background scalar field is of the order of the Hubble time and there exists
a set of orthonormal basis vectors, $e^\mathrm{a}_i$ ($i=1,2,\cdots,D$),
in the field space that remain approximately constant in time.
In other words, we assume that all the frequencies associated with quantum
fluctuations of the scalar field are much larger than $H$ and their
components with respect to the basis $e^\mathrm{a}_i$
can be regarded as mutually independent.

Under the above assumption, we quantize the scalar fields as
\begin{equation}
\delta\phi_\mathrm{f}^\mathbf{a}(\mathbf{k},t) = \sum_{i=1}^D \left[ a_i(\mathbf{k}) \,
 \delta\phi_i^\mathbf{a}(k,t) + a_i^\dagger(-\mathbf{k}) \,
 {\delta\phi_i^\mathbf{a}}^\dagger(k,t) \right] ~,
\end{equation}
where $a_i(\mathbf{k})$ and $a_i^\dag(\mathbf{k})$ are
annihilation and creation operators, and
the scalar field mode functions $\delta\phi_i^\mathbf{a}(k,t)$
are normalized as
\begin{eqnarray}
h_\mathbf{ab} \left[ \dot{\delta\phi_i^\mathbf{a}}(k,t) \,
 \delta\phi_j^\mathbf{b}(k,t) - \delta\phi_i^\mathbf{a}(k,t) \,
 \dot{\delta\phi_j^\mathbf{b}}(k,t) \right] & = & 0 ~,
\\
h_\mathbf{ab} \left[ \dot{\delta\phi_i^\mathbf{a}}^\dagger(k,t) \,
 \delta\phi_j^\mathbf{b}(k,t) - {\delta\phi_i^\mathbf{a}}^\dagger(k,t) \,
 \dot{\delta\phi_j^\mathbf{b}}(k,t) \right] & = & 2i q^3 \delta_{ij} ~.
\end{eqnarray}
The mode functions have the asymptotic behavior
\begin{equation}
\lim_{x \rightarrow \infty} \delta\phi_i^\mathbf{a} = q e^{ix} e_i^\mathbf{a},
\label{phiasymp}
\end{equation}
\begin{equation}
\lim_{x \rightarrow \infty} \dot{\delta\phi_i^\mathbf{a}}
= - i q^2 e^{ix} e_i^\mathbf{a},
\label{dotphiasymp}
\end{equation}
with
\begin{equation}\label{basis}
h_\mathbf{ab} e^\mathbf{a}_i e^\mathbf{b}_j = \delta_{ij}
\comma
\sum_{i=1}^D e^\mathbf{a}_i e^\mathbf{b}_i = h^\mathbf{ab} ~.
\end{equation}
Then the creation and annihilation operators obey the quantization conditions
\begin{equation}
\left[ a_i(\mathbf{k}) , a_j^\dagger(\mathbf{l}) \right] = \frac{1}{2k^3} \delta_{ij} \,
 \delta(\mathbf{k}-\mathbf{l}) ~,
\end{equation}
\begin{equation}
\left[ a_i(\mathbf{k}) , a_j(\mathbf{l}) \right] = 0
 = \left[ a_i^\dagger(\mathbf{k}) , a_j^\dagger(\mathbf{l}) \right] ~.
\end{equation}

Given the quantized scalar field perturbations $\delta\phi^\mathbf{a}_\mathrm{f}$,
the quantized ${\delta N}_\mathrm{f}$ is expressed as
\begin{equation}
{\delta N}_\mathrm{f}(\mathbf{k},x)
= \sum_{i=1}^D \left[ a_i (\mathbf{k}) \, {\delta N}_i(k,x)
 + a_i^\dagger(-\mathbf{k}) \, {\delta N}_i^*(k,x) \right] ~,
\end{equation}
where, from Eq.~(\ref{deltaNbb}), ${\delta N}_i(k,x)$ is given by
\begin{equation}
{\delta N}_i = \mathbb{N}_{\phi^\mathbf{a}} \delta\phi_i^\mathbf{a}
 + \mathbb{N}_{\dot\phi^\mathbf{a}} \dot{\delta\phi_i^\mathbf{a}} ~.
\end{equation}
Eq.~(\ref{deltaNeomx}) translates to
\begin{equation}\label{deltaNieom}
{\delta N}_i'' - \frac{4}{x} {\delta N}_i' + {\delta N}_i
 = \frac{2}{x} \frac{\Pi'}{\Pi} {\delta N}_i'
 - 2 \dot\mathbb{N}_{\dot\phi^\mathbf{a}} \delta\phi_{i\bot}^\mathbf{a} ~,
\end{equation}
with the asymptotic behavior
\begin{equation}
\lim_{x \rightarrow \infty} {\delta N}_i
 = - i q^2 e^{ix} \mathbb{N}_{\dot\phi^\mathbf{a}} e_i^\mathbf{a}
\label{Niasymp}
\end{equation}
following from Eqs.~(\ref{phiasymp}) and~(\ref{dotphiasymp}).

The power spectrum is given by
\begin{equation}
\frac{2\pi^2}{k^3} P(k) \, \delta(\mathbf{k}-\mathbf{l})
= \lim_{x \rightarrow 0} \left\langle {\delta N}_\mathrm{f}(\mathbf{k},x) \,
 {\delta N}_\mathrm{f}^\dagger(\mathbf{l},x) \right\rangle ~.
\end{equation}
Therefore
\begin{equation}
P(k) = \frac{1}{(2\pi)^2} \lim_{x \rightarrow 0}
 \sum_{i=1}^D \left| {\delta N}_i(k,x) \right|^2 ~.
\label{Pformula}
\end{equation}
Thus what we have to do is to solve Eq.~(\ref{deltaNieom}) for $\delta N_i$
with the asymptotic boundary condition~(\ref{Niasymp}).

\subsection{Zeroth order}

To zeroth order, we neglect the terms on the right hand side of Eq.~(\ref{deltaNieom}),
as they generate deviations from scale invariance, leaving
\begin{equation}\label{deltaN0eom}
{{\delta N}_i^{(0)}}'' - \frac{4}{x} {{\delta N}_i^{(0)}}' + {\delta N}_i^{(0)} = 0 ~.
\end{equation}
This equation has mode functions
\begin{equation}
n_0(x) = \left( 1 - i x - \frac{1}{3} x^2 \right) e^{ix} ~.
\end{equation}
Properties of $n_0(x)$ are given in Appendix~\ref{n0}.
The solution for ${\delta N}_i^{(0)}$ with the asymptotic behavior (\ref{Niasymp}) is
\begin{equation}
{\delta N}_i^{(0)}(k,x) = 2\pi i \, \Pi_i(\infty) \, n_0(x) ~,
\end{equation}
where
\begin{equation}\label{Pii}
\Pi_i \equiv \frac{3 \mathbb{N}_{\dot\phi^\mathbf{a}} e^\mathbf{a}_i}{2\pi a^2 \xi^2} ~.
\end{equation}
Note that
\begin{equation}
\sum_{i=1}^D\Pi_i^2 = \Pi^2 ~,
\end{equation}
where $\Pi^2$ was defined in Eq.~(\ref{Pi2}).
Therefore the zeroth order power spectrum is
\begin{equation}
P(k) = \Pi^2(\infty) ~.
\end{equation}
Note that at this order $\Pi^2$ is regarded as constant and so the evaluation
 at $x \rightarrow \infty$ can be dropped.

\subsection{First order}

The Green's function solution of Eq.~(\ref{deltaNieom}) is
\begin{eqnarray}
{\delta N}_i(k,x)
& = & {\delta N}_i^{(0)}(k,x) + 18 \int_x^\infty \frac{du}{u^4}
 \left\{ \mathrm{Re} \left[ n_0(x) \right] \mathrm{Im} \left[ n_0(u) \right]
 - \, \mathrm{Im} \left[ n_0(x) \right] \mathrm{Re} \left[ n_0(u) \right] \right\}
\nonumber \\ && \times
\left[ \frac{1}{u} \frac{\Pi'}{\Pi} {\delta N}_i'(k,u)
 - \dot\mathbb{N}_{\dot\phi^\mathbf{a}}
 \delta\phi_{i\bot}^\mathbf{a}(k,u) \right] ~.
\end{eqnarray}
Taking the limit $x \rightarrow 0$ and using Eq.~(\ref{n0x0}), we get
\begin{eqnarray}
\lim_{x \rightarrow 0} {\delta N}_i(k,x)
 & = & \lim_{x \rightarrow 0} {\delta N}_i^{(0)}(k,x)
 + 18 \lim_{x \rightarrow 0} \int_x^\infty \frac{du}{u^4}
 \left\{ \mathrm{Im} \left[ n_0(u) \right]
 - \frac{1}{45} x^5 \mathrm{Re} \left[ n_0(u) \right] \right\}
\nonumber \\ && \times
\left[ \frac{1}{u} \frac{\Pi'}{\Pi} {\delta N}_i'(k,u)
 - \dot\mathbb{N}_{\dot\phi^\mathbf{a}}
 \delta\phi_{i\bot}^\mathbf{a}(k,u) \right] ~.
\end{eqnarray}
As long as $x^2 \Pi'/\Pi$ and
$x^2 \dot\mathbb{N}_{\dot\phi^\mathbf{a}} \, \delta\phi_{i\bot}^\mathbf{a}$
become negligible as $x \rightarrow 0$ before $\Pi'/\Pi$ and
$\dot\mathbb{N}_{\dot\phi^\mathbf{a}} \, \delta\phi_{i\bot}^\mathbf{a}$
stop being small, \mbox{i.e.} before our approximation scheme breaks down,
then we can complete the limit $x \rightarrow 0$ without needing to apply distinct superhorizon methods \cite{superhorizon,SSTY}.
Assuming this and completing the limit we get
\begin{equation}
\lim_{x \rightarrow 0} {\delta N}_i(k,x) = {\delta N}_i^{(0)}(k,0)
 + 18 \int_0^\infty \frac{du}{u^4} \, \mathrm{Im} \left[ n_0(u) \right]
 \left[ \frac{1}{u} \frac{\Pi'}{\Pi} {\delta N}_i'(k,u)
 - \dot\mathbb{N}_{\dot\phi^\mathbf{a}}
 \delta\phi_{i\bot}^\mathbf{a}(k,u) \right] ~.
\end{equation}
Substituting this into Eq.~(\ref{Pformula}), we obtain
\begin{equation}
(2\pi)^2 P(k) = \sum_{i=1}^D
 \left| {\delta N}_i^{(0)}(k,0) + 18 \int_0^\infty \frac{du}{u^4} \,
 \mathrm{Im} \left[ n_0(u) \right] \left[ \frac{1}{u} \frac{\Pi'}{\Pi}
 {\delta N}_i'(k,u) -  \dot\mathbb{N}_{\dot\phi^\mathbf{a}}
 \delta\phi_{i\bot}^\mathbf{a}(k,u) \right] \right|^2 ~.
\end{equation}
Keeping all the potentially leading order corrections from each term
on the right hand side of Eq.~(\ref{deltaNieom}), we have
\begin{eqnarray}
\lefteqn{
(2\pi)^2 P(k) \simeq \sum_{i=1}^D \left| {\delta N}_i^{(0)}(k,0) \right|^2
} \nonumber \\ && \mbox{}
 + 18 \sum_{i=1}^D {{\delta N}_i^{(0)}}^*(k,0) \int_0^\infty \frac{du}{u^4} \,
 \mathrm{Im} \left[ n_0(u) \right] \left[ \frac{1}{u} \frac{\Pi'}{\Pi}
 {{\delta N}_i^{(0)}}'(k,u) - \dot\mathbb{N}_{\dot\phi^\mathbf{a}}
 \delta\phi_{i\bot}^\mathbf{a}(k,u) \right] + \mbox{c.c.}
\nonumber \\ && \mbox{}
+ \sum_{i=1}^D \left| 18 \int_0^\infty \frac{du}{u^4} \,
 \mathrm{Im} \left[ n_0(u) \right]
 \dot\mathbb{N}_{\dot\phi^\mathbf{a}}
 \delta\phi_{i\bot}^\mathbf{a}(k,u) \right|^2 ~.
\end{eqnarray}
Defining
\begin{equation}
z_\parallel(x) \equiv \frac{18}{x^4} \,
 \mathrm{Im}\left[n_0(x)\right] \mathrm{Re}\left[n_0'(x)\right]
\end{equation}
and
\begin{equation}
z_\bot(x) \equiv \frac{9}{x^3} \mathrm{Im}\left[n_0(x)\right]
\end{equation}
with properties given in Appendix~\ref{n0}, we have
\begin{eqnarray}
P(k) & = & \Pi^2(\infty)
+ 2 \, \Pi^2(\infty) \int_0^\infty \frac{d\xi}{\xi} \, z_\parallel(k\xi) \,
 \frac{\Pi'}{\Pi}(\xi)
\nonumber \\ && \mbox{}
- \frac{2}{\pi} \sum_{i=1}^D \Pi_i(\infty)
 \int_0^\infty \frac{d\xi}{\xi} \, z_\bot(k\xi) \,
  \dot\mathbb{N}_{\dot\phi^\mathbf{a}}(\xi) \,
 \mathrm{Im}\left[\delta\phi_{i\bot}^\mathbf{a}(k,\xi)\right]
\nonumber \\ && \mbox{} \label{Pfull}
+ \frac{1}{\pi^2} \sum_{i=1}^D \left| \int^\infty_0 \frac{d\xi}{\xi} \,
 z_\bot(k\xi) \, \dot\mathbb{N}_{\dot\phi^\mathbf{a}}(\xi) \,
 \delta\phi_{i\bot}^\mathbf{a}(k,\xi) \right|^2 ~.
\end{eqnarray}

For scale invariance of the spectrum, we require $\Pi^2$ to be approximately constant.
A reasonable extrapolation of this is to assume that
$\mathbb{N}_{\dot\phi^\mathbf{a}}$ is approximately constant, and hence
$\dot\mathbb{N}_{\dot\phi^\mathbf{a}}$ is small.
This would naively make the third term in Eq.~(\ref{Pfull}) first order
and the last term second order.
However, from Eqs.~(\ref{PQdeltaphi}) and~(\ref{Pii}),
\begin{equation}
\sum_{i=1}^D \Pi_i \, \delta\phi_{i\bot}^\mathbf{a}
= \frac{3}{2\pi a^2 \xi^2} \sum_{i=1}^D \mathbb{N}_{\dot\phi^\mathbf{c}}
 e^\mathbf{c}_i Q^\mathbf{a}_\mathbf{b} \delta\phi_i^\mathbf{b} ~.
\end{equation}
Using Eqs.~(\ref{phiasymp}) and~(\ref{basis}),
we see that this vanishes in the limit $\xi \rightarrow \infty$.
If $\delta\phi_\parallel^\mathbf{a}$ and $\delta\phi_\bot^\mathbf{a}$ only weakly mix,
then the third term in Eq.~(\ref{Pfull}) will be further suppressed
making it naively second order.
Then the last two terms in Eq.~(\ref{Pfull}) could both be regarded as second order,
though as now at first order we have no direct constraints on $\delta\phi_\bot^\mathbf{a}$
one should be somewhat cautious about this.

If we neglect the $\delta\phi_\bot^\mathbf{a}$ terms, Eq.~(\ref{Pfull}) reduces to
\begin{eqnarray}
\ln P(k) & = & \ln\Pi^2(\infty) + 2 \int_0^\infty \frac{d\xi}{\xi} \,
 z_\parallel(k\xi) \, \frac{\Pi'}{\Pi}(\xi)
\\
& = & \ln\Pi^2(\infty) + \int_0^\infty \frac{d\xi}{\xi}
 \left[ - k\xi \, z_\parallel^\prime(k\xi) \right] \ln \frac{\Pi^2(\xi)}{\Pi^2(\infty)}
\\
& = & \int_0^\infty \frac{d\xi}{\xi} \left[ - k \xi \, W'(k\xi) \right]
\left[ \ln\Pi^2 + \frac{2}{3} \frac{\Pi'}{\Pi} \right](\xi) ~,
\label{PW}
\end{eqnarray}
where the definition and properties of $W(x)$ are given in Appendix~\ref{n0}.
This has a form equivalent to the single component case \cite{Scott,gsr,JY},
though with a different interpretation of the inflationary source
$\ln\Pi^2 + 2\Pi' / 3\Pi$.
\footnote{See Appendix~\ref{AppSingle} for the equivalence between
the single field reduction of Eq.~(\ref{PW}) and the single field
general slow-roll formula of Refs.~\cite{Scott,gsr,JY}.}

\subsection{Inverse formula}

In the special case when the spectrum has the form of Eq.~(\ref{PW}),
it is possible to invert the spectrum to obtain the function $\Pi^2$,
in the same way as was done in the single field case~\cite{Minu}.

Let us introduce the function
\begin{equation}
s(x) \equiv \frac{2}{\pi} \left[ \frac{3}{x^3} - \frac{3\cos(2x)}{x^3}
 - \frac{6\sin(2x)}{x^2} + \frac{2}{x} + \frac{4\cos(2x)}{x} + \sin(2x) \right] ~,
\end{equation}
which has the window property
\begin{equation}
\int_0^\infty \frac{dx}{x} \, s(x) = 1 ~,
\label{sint}
\end{equation}
the asymptotic behavior
\begin{equation}
\lim_{x \rightarrow 0} s(x) = \frac{4}{45\pi} x^5 + \mathcal{O}(x^7) ~,
\end{equation}
and the inverse properties
\begin{equation}
\int_0^\infty \frac{d\xi}{\xi} \, s(k\xi) \, z_\parallel(l\xi) = - \theta(l-k)
\label{szpara}
\end{equation}
or
\begin{equation}
\int_0^\infty \frac{d\xi}{\xi} \, s(k\xi) \,
 W(l\xi) = - \theta(l-k) - \frac{l^3}{k^3} \, \theta(k-l) ~.
\end{equation}
Then, the function $s(x)$ allows us to invert Eq.~(\ref{PW}) to get
\begin{equation}\label{inverse}
\ln\Pi^2(\xi) = \int_0^\infty \frac{dk}{k} \, s(k\xi) \ln P(k) ~.
\end{equation}

\section{Summary and Conclusions}
\label{sac}

We have developed a new $\delta N$ formalism for multi-component inflation
that can be applied in the most general situations.
We have introduced a perturbation variable $\delta N$
whose form closely resembles the deviation of the $e$-folding number of expansion
between two adjacent background trajectories in phase space,
and whose evolution can be solved from sub-horizon to super-horizon scales.
We have shown, though not rigorously yet, that
$\delta N(t_\mathrm{ini}) = \delta\mathcal{N}(t_\mathrm{fin},t_\mathrm{ini})$
on super-horizon scales, where $\delta\mathcal{N}(t_\mathrm{fin},t_\mathrm{ini})$
describes the perturbation of the $e$-folding number of expansion
between a time $t_\mathrm{ini}$ during inflation
and the final comoving hypersurface $t_\mathrm{fin}$,
on which the curvature perturbation $\mathcal{R}_\mathrm{c}$
has achieved its final late time constant value.

Taking the flat slicing during inflation, denoted by subscript f,
we have derived a simple form for the evolution equation for $\delta N_\mathrm{f}$.
This equation contains a source term
that involves the scalar field perturbations explicitly.
However, this source term is of first order in the deviations from scale invariance,
and furthermore, its contribution to the spectrum of the curvature perturbation,
which we have computed to leading order, can be expected to be
of second order in the deviations from scale invariance.
In the case when this source term can be neglected,
which includes the case of conventional slow-roll inflation,
we have obtained the power spectrum for $\mathcal{R}_\mathrm{c}$
in a closed form in terms of functions determined by the background dynamics.
In this case, we have found that the spectral formula is invertible,
\mbox{i.e.} we have derived a formula that expresses
the information of the inflationary dynamics directly
in terms of the final curvature perturbation spectrum.

The formula for the spectrum of $\mathcal{R}_\mathrm{c}$ derived in this paper
can be applied to a very wide class of models.
Nevertheless, it will be useful to extend our formula to second order for greater accuracy and to include the possibility of super-horizon contributions \cite{superhorizon}.
Considering special cases, such as the standard slow-roll limit, may help to make our formalism more accessible, and application to physically interesting models will also be useful.
In addition, as mentioned above, we have not given a rigorous proof of the equivalence between $\delta N$ and $\delta\mathcal{N}$ on super-horizon scales.
We plan to explore these issues in a forthcoming paper~\cite{SSTY}.

\subsection*{Acknowledgements}
This work was supported in part by
ARCSEC funded by the Korea Science and Engineering Foundation and the
Korean Ministry of Science,
the Korea Research Foundation grant KRF PBRG 2002-070-C00022,
Brain Korea 21,
an Erskine Fellowship of the University of Canterbury,
and JSPS Grant-in-Aid for Scientific Research, Nos.~14102004,
16740165 and 17340075.

\appendix
\section{Covariant partial derivatives with respect to ($\bm{\phi,\dot\phi}$) and ($\bm{\phi,\pi}$)}
\label{cpd}

First we define the covariant partial derivatives in the phase space $(\phi,\dot\phi)$.
For a scalar function $F(\phi,\dot\phi)$
and phase space coordinates $(\phi^\alpha,\dot\phi^\alpha)$, we have
\begin{eqnarray}\label{coordpd}
\dot{F} = F_{\phi^\alpha} \dot\phi^\alpha + F_{\dot\phi^\alpha} \ddot\phi^\alpha ~,
\end{eqnarray}
where $F_{\phi^\alpha}$ and $F_{\dot \phi^\alpha}$ are the partial derivatives with respect to the coordinates $\phi^\alpha$ and $\dot\phi^\alpha$,
and $\dot\phi^{\alpha}$ and $\ddot\phi^{\alpha}$ are the time derivatives of the coordinates $\phi^\alpha$ and $\dot\phi^{\alpha}$.
However, we wish to write all equations in a geometric manner.
In order not to change the form of Eq.~(\ref{coordpd}), we should define the covariant partial derivatives as
\begin{equation}
F_{\phi^\mathbf{a}} \equiv F_{\phi^\alpha} e^\alpha_\mathbf{a}
 + F_{\dot\phi^\alpha} \dot{e^\alpha_\mathbf{a}}
\comma
F_{\dot\phi^\mathbf{a}} \equiv F_{\dot\phi^\alpha} e^\alpha_\mathbf{a} ~,
\end{equation}
\begin{equation}
\left(e_\alpha^\mathbf{a}\right)_{\phi^\mathbf{b}}
 \equiv \nabla_\mathbf{b} e_\alpha^\mathbf{a}
\comma
\left(e_\alpha^\mathbf{a}\right)_{\dot\phi^\mathbf{b}} \equiv 0 ~,
\end{equation}
where the $e_\alpha^\mathbf{a}$ with $\alpha = 1,\ldots,D$ are a complete set of basis vectors in the $D$-dimensional field space.
Then we have
\begin{equation}
\dot{F} = F_{\phi^\mathbf{a}} \dot\phi^\mathbf{a}
 + F_{\dot\phi^\mathbf{a}} \ddot\phi^\mathbf{a} ~,
\label{dotf}
\end{equation}
\begin{equation}
\dot\phi^\mathbf{a}_{\phi^\mathbf{b}} = 0
\comma
\dot\phi^\mathbf{a}_{\dot\phi^\mathbf{b}} = \delta^\mathbf{a}_\mathbf{b} ~,
\end{equation}
\begin{equation}
F_{\phi^\mathbf{b}\phi^\mathbf{a}} - F_{\phi^\mathbf{a}\phi^\mathbf{b}}
 = F_{\dot\phi^\mathbf{c}} {R^\mathbf{c}}_\mathbf{bda} \dot\phi^\mathbf{d} ~,
\end{equation}
\begin{equation}
F_{\dot\phi^\mathbf{b}\phi^\mathbf{a}} - F_{\phi^\mathbf{a}\dot\phi^\mathbf{b}}
= F_{\dot\phi^\mathbf{b}\dot\phi^\mathbf{a}} - F_{\dot\phi^\mathbf{a}\dot\phi^\mathbf{b}}
= 0 ~,
\end{equation}
where ${R_\mathbf{abc}}^\delta \equiv \left( \nabla_\mathbf{a} \nabla_\mathbf{b}
 - \nabla_\mathbf{b} \nabla_\mathbf{a} \right) e^\delta_\mathbf{c}$
 is the curvature tensor on the scalar field space.

Similarly, we may define the covariant partial derivatives
in the phase space $(\phi,\pi)$, where
\begin{equation}
\pi^\mathbf{a} \equiv \frac{\dot\phi^\mathbf{a}}{H} ~.
\end{equation}
For a scalar function $G(\phi,\pi)$, we define
\begin{equation}
G_{\phi^\mathbf{a}} \equiv G_{\phi^\alpha} e^\alpha_\mathbf{a}
 + G_{\pi^\alpha} \frac{\dot{e^\alpha_\mathbf{a}}}{H}
\comma
G_{\pi^\mathbf{a}} \equiv G_{\pi^\alpha} e^\alpha_\mathbf{a} ~,
\end{equation}
\begin{equation}
\left(e_\alpha^\mathbf{a}\right)_{\phi^\mathbf{b}}
 \equiv \nabla_\mathbf{b} e_\alpha^\mathbf{a}
\comma
\left(e_\alpha^\mathbf{a}\right)_{\pi^\mathbf{b}} \equiv 0 ~.
\end{equation}
Then we have
\begin{equation}
\frac{\dot{G}}{H} = G_{\phi^\mathbf{a}} \pi^\mathbf{a} + G_{\pi^\mathbf{a}} \frac{\dot\pi^\mathbf{a}}{H} ~,
\end{equation}
\begin{equation}
\pi^\mathbf{a}_{\phi^\mathbf{b}} = 0
\comma
\pi^\mathbf{a}_{\pi^\mathbf{b}} = \delta^\mathbf{a}_\mathbf{b} ~,
\end{equation}
\begin{equation}
G_{\phi^\mathbf{b}\phi^\mathbf{a}} - G_{\phi^\mathbf{a}\phi^\mathbf{b}}
 = G_{\pi^\mathbf{c}} {R^\mathbf{c}}_\mathbf{bda} \pi^\mathbf{d} ~,
\end{equation}
\begin{equation}
G_{\pi^\mathbf{b}\phi^\mathbf{a}} - G_{\phi^\mathbf{a}\pi^\mathbf{b}} =
G_{\pi^\mathbf{b}\pi^\mathbf{a}} - G_{\pi^\mathbf{a}\pi^\mathbf{b}} = 0 ~.
\end{equation}

The relationship between these partial derivatives is given by
\begin{eqnarray}
\left( \frac{\partial Q}{\partial\phi^\mathbf{a}} \right)_{\dot\phi}
& = & \left( \frac{\partial Q}{\partial\phi^\mathbf{a}} \right)_\pi
+ \left( \frac{\partial Q}{\partial\pi^\mathbf{b}} \right)_\phi
 \left( \frac{\partial\pi^\mathbf{b}}{\partial\phi^\mathbf{a}} \right)_{\dot\phi}
\nonumber\\
& = & \left( \frac{\partial Q}{\partial\phi^\mathbf{a}} \right)_\pi
- \frac{1}{H} \left( \frac{\partial Q}{\partial\pi^\mathbf{b}} \right)_\phi
 \pi^\mathbf{b} \left( \frac{\partial H}{\partial\phi^\mathbf{a}} \right)_{\dot\phi}
\nonumber\\
& = & \left( \frac{\partial Q}{\partial\phi^\mathbf{a}} \right)_\pi
- \frac{1}{6 H^2} \left( \frac{\partial Q}{\partial\pi^\mathbf{b}} \right)_\phi
 \pi^\mathbf{b} V_{\phi^\mathbf{a}} ~,
\end{eqnarray}
\begin{eqnarray}
\left( \frac{\partial Q}{\partial\dot\phi^\mathbf{a}} \right)_\phi
& = & \left( \frac{\partial Q}{\partial\pi^\mathbf{b}} \right)_\phi
 \left( \frac{\partial\pi^\mathbf{b}}{\partial\dot\phi^\mathbf{a}} \right)_\phi
\nonumber\\
& = & \frac{1}{H} \left( \frac{\partial Q}{\partial\pi^\mathbf{b}} \right)_\phi
 \left[ \delta_\mathbf{a}^\mathbf{b} - \pi^\mathbf{b}
 \left( \frac{\partial H}{\partial\dot\phi^\mathbf{a}} \right)_\phi \right]
\nonumber\\
& = & \frac{1}{H} \left( \frac{\partial Q}{\partial\pi^\mathbf{b}} \right)_\phi
 \left[ \delta_\mathbf{a}^\mathbf{b} - \frac{1}{6} \pi^\mathbf{b} \pi_\mathbf{a} \right]
\end{eqnarray}
and
\begin{eqnarray}
\left( \frac{\partial Q}{\partial\dot\phi^\mathbf{a}} \right)_\phi \dot\phi^\mathbf{a}
= \left( 1 + \frac{1}{3} \frac{\dot{H}}{H^2} \right)
 \left( \frac{\partial Q}{\partial\pi^\mathbf{a}} \right)_\phi \pi^\mathbf{a} ~.
\label{dotphipirel}
\end{eqnarray}

In particular
\begin{eqnarray}\label{dotphitopi1}
\delta Q & \equiv &
\left( \frac{\partial Q}{\partial\phi^\mathbf{a}} \right)_{\dot\phi}
 \delta\phi^\mathbf{a}
+ \left( \frac{\partial Q}{\partial\dot\phi^\mathbf{a}} \right)_\phi
 \left( \dot{\delta\phi^\mathbf{a}} - \dot\phi^\mathbf{a} A \right)
\\ \label{dotphitopi2}
& = & \left( \frac{\partial Q}{\partial\phi^\mathbf{a}} \right)_\pi
 \delta\phi^\mathbf{a}
+ \left( \frac{\partial Q}{\partial\pi^\mathbf{a}} \right)_\phi
\left\{ \delta\pi^\mathbf{a} - \pi^\mathbf{a} \left[ \frac{1}{H} \dot\mathcal{R}
 - \frac{q^2}{3H} \left( \mathcal{S} - \frac{1}{H} \mathcal{R} \right) \right] \right\}
\\ \label{dotphitopi3}
& = & \left( \frac{\partial Q}{\partial\phi^\mathbf{a}} \right)_\pi
 \delta\phi^\mathbf{a}
+ \left( \frac{\partial Q}{\partial\pi^\mathbf{a}} \right)_\phi
 \left\{ \delta\pi^\mathbf{a}
 - \pi^\mathbf{a} \left[ A + \mathcal{K} + \frac{q^2}{3H^2} \mathcal{R} \right] \right\} ~,
\end{eqnarray}
where
\begin{equation}
\delta\pi^\mathbf{a} \equiv \frac{\dot{\delta\phi^\mathbf{a}}}{H} ~.
\label{deltapidef}
\end{equation}
As noted in Section~\ref{secdeltaN}, we may interpret
\begin{equation}
\dot{\delta\phi^\mathbf{a}} - \dot\phi^\mathbf{a} A
 = \delta \left( \frac{d\phi^\mathbf{a}}{d\tau} \right) ~,
\end{equation}
which makes the form of $\delta Q$ in Eq.~(\ref{dotphitopi1})
identical to that of the perturbation of the background quantity.
Furthermore, noting that
\begin{equation}\label{deltaH}
\left( \frac{\partial H}{\partial\phi^\mathbf{a}} \right)_{\dot\phi}
 \delta\phi^\mathbf{a}
 + \left( \frac{\partial H}{\partial\dot\phi^\mathbf{a}} \right)_\phi
 \left( \dot{\delta\phi^\mathbf{a}} - \dot\phi^\mathbf{a} A \right)
 = H \mathcal{K} + \frac{q^2}{3H} \mathcal{R} ~,
\end{equation}
we find it is also possible to interpret
\begin{equation}\label{deltaPi}
\delta\pi^\mathbf{a} - \pi^\mathbf{a} \left[ A + \mathcal{K}
 + \frac{q^2}{3H^2} \mathcal{R} \right]
 = \delta \left( \frac{d\phi^\mathbf{a}}{d\mathcal{N}} \right) ~,
\end{equation}
at least on super-horizon scales or flat hypersurfaces.

\section{Properties of $\bm{n_0(x)}$, $\bm{z_\parallel(x)}$, $\bm{z_\bot(x)}$,
 $\bm{W(x)}$ and $\bm{s(x)}$}
\label{n0}

Here we list some mathematical properties of the functions used in Section~\ref{mfgsrps}.

\noindent
Definitions:
\begin{equation}
n_0(x) = \left( 1 - ix - \frac{1}{3}x^2 \right) e^{ix} ~,
\end{equation}
\begin{equation}
z_\parallel(x)
= \frac{3\sin(2x)}{x^3} - \frac{6\cos(2x)}{x^2} - \frac{4\sin(2x)}{x} + \cos(2x) - 1 ~,
\end{equation}
\begin{equation}
z_\bot(x) = \frac{9\sin x}{x^3} - \frac{9\cos x}{x^2} - \frac{3\sin x}{x} ~,
\end{equation}
\begin{equation}
W(x) = \frac{3\sin(2x)}{2x^3} - \frac{3\cos(2x)}{x^2} - \frac{3\sin(2x)}{2x} - 1 ~,
\label{Wdef}
\end{equation}
\begin{equation}
s(x) = \frac{2}{\pi} \left[ \frac{3}{x^3} - \frac{3\cos(2x)}{x^3}
 - \frac{6\sin(2x)}{x^2} + \frac{2}{x} + \frac{4\cos(2x)}{x} + \sin(2x) \right] ~.
\end{equation}
Relations:
\begin{equation}
n_0'' - \frac{4}{x} n_0' + n_0 = 0 ~,
\end{equation}
\begin{equation}
\mathrm{Re}\left[n_0(x)\right] \mathrm{Im}\left[n_0(x)\right]'
 - \mathrm{Re}\left[n_0(x)\right]' \mathrm{Im}\left[n_0(x)\right] = \frac{1}{9} x^4 ~,
\end{equation}
\begin{equation}
z_\parallel(x) = \frac{18}{x^4} \mathrm{Im}\left[n_0(x)\right]
 \mathrm{Re}\left[n_0^\prime(x)\right]
= W(x) - \frac{x}{3} \, W'(x) ~,
\end{equation}
\begin{equation}
z_\bot(x) = \frac{9}{x^3} \mathrm{Im}\left[n_0(x)\right] ~.
\end{equation}
Asymptotics:
\begin{equation}\label{n0x0}
\lim_{x \rightarrow 0} n_0(x)
 = 1 + \frac{1}{6} x^2 + \mathcal{O}(x^4) + \frac{i}{45} x^5 + \mathcal{O}(ix^7) ~,
\end{equation}
\begin{equation}
\lim_{x \rightarrow 0} z_\parallel(x) = \frac{2}{15} x^2 + \mathcal{O}(x^4) ~,
\end{equation}
\begin{equation}
\lim_{x \rightarrow 0} z_\bot(x) = \frac{1}{5} x^2 + \mathcal{O}(x^4) ~,
\end{equation}
\begin{equation}
\lim_{x \rightarrow 0} W(x) = \frac{2}{5} x^2 + \mathcal{O}(x^4) ~,
\end{equation}
\begin{equation}
\lim_{x \rightarrow 0} s(x) = \frac{4}{45\pi} x^5 + \mathcal{O}(x^7) ~.
\end{equation}
Window property:
\begin{equation}
\int_0^\infty \frac{dx}{x} z_\bot(x) = 1 ~,
\end{equation}
\begin{equation}
\int_0^\infty \frac{dx}{x} \left[-x\,W'(x)\right] = 1 ~,
\end{equation}
\begin{equation}
\int_0^\infty \frac{dx}{x} \, s(x) = 1 ~.
\end{equation}
Degeneracy:
\begin{equation}
\int_0^\infty \frac{dx}{x} \frac{1}{x} \, z_\parallel(x)
= \int_0^\infty \frac{dx}{x} \frac{1}{x} \left[- x z_\parallel'(x) \right]
= 0 ~,
\end{equation}
\begin{equation}
\int_0^\infty \frac{dx}{x} \frac{1}{x} \, W(x)
= \int_0^\infty \frac{dx}{x} \frac{1}{x} \left[-x\,W'(x)\right] = 0 ~.
\end{equation}
Inversion:
\begin{equation}
\int_0^\infty \frac{dk}{k} \, s(k\zeta) \, z_\parallel(k\xi) = - \theta(\xi-\zeta) ~,
\end{equation}
\begin{equation}
\int_0^\infty \frac{dk}{k} \, s(k\zeta) \, W(k\xi)
 = - \theta(\xi-\zeta) - \frac{\xi^3}{\zeta^3} \, \theta(\zeta-\xi) ~.
\end{equation}
where $\theta(x) = 0$ for $x < 0$ and $\theta(x) = 1$ for $x > 0$.
See Ref.~\cite{gsr} for a graph of the window function $- x \, W'(x)$.

\section{Connection with single field formula}
\label{AppSingle}

The single field general slow-roll formula for the curvature perturbation spectrum is \cite{Scott,gsr,JY}
\begin{equation}
\ln P = \int_0^\infty \frac{d\xi}{\xi} \left[ - k \xi \, W'(k\xi) \right]
 \left[ \ln \frac{1}{f^2} + \frac{2}{3} \frac{f'}{f} \right] ~,
\end{equation}
where
\begin{equation}
f(\ln\xi) = \frac{2\pi a \xi \dot\phi}{H} ~.
\end{equation}
This is to be compared with the single field reduction of Eq.~(\ref{PW}),
\begin{equation}
\ln P = \int_0^\infty \frac{d\xi}{\xi} \left[ - k \xi \, W'(k\xi) \right]
 \left[ \ln\Pi^2 + \frac{2}{3} \frac{\Pi'}{\Pi} \right] ~,
\end{equation}
where
\begin{equation}
\Pi^2 = \left( \frac{3 \mathbb{N}_{\dot\phi}}{2\pi a^2 \xi^2} \right)^2
\end{equation}
and
\begin{equation}
\mathbb{N}_{\dot\phi}
 = N_{\dot\phi} - \frac{a^3\dot\phi}{6H} \int^t_{t_\mathrm{fin}}\frac{dt}{a^3} ~.
\end{equation}

The single field formula can be inverted to obtain information about
 the inflaton dynamics from the primordial spectrum \cite{Minu},
\begin{equation}
\ln\frac{1}{f^2} = \int_0^\infty \frac{dk}{k} \, m(k\xi) \ln P ~.
\end{equation}
This may be compared with the single field reduction of Eq.~(\ref{inverse}),
\begin{equation}
\ln \Pi^2 = \int_0^\infty \frac{dk}{k} \, s(k\xi) \ln P ~,
\end{equation}
where $m(x)$ and $s(x)$ are related by
\begin{equation}
m(x) - \frac{x}{3} \, m'(x) = s(x) + \frac{x}{3} \, s'(x) ~.
\end{equation}

Thus the single field formulae and the single field reductions
of the multi-field formulae are equivalent if
\begin{equation}
\ln\Pi^2 + \frac{2}{3} \frac{\Pi'}{\Pi}
 \simeq \ln \frac{1}{f^2} + \frac{2}{3} \frac{f'}{f} ~.
\end{equation}
To check the above correspondence explicitly, we first
note that in the single field case
Eq.~(\ref{Ndotphidotphi}) reduces to
\begin{equation}
\left( \frac{H N_{\dot\phi}}{a^3 \dot\phi} \right)^\centerdot
 = - \left( \frac{H^2}{\dot\phi^2} - \frac{1}{6} \right) \frac{1}{a^3} ~.
\end{equation}
Integrating this in time, we have
\begin{equation}
N_{\dot\phi} = - \frac{a^3 \dot\phi}{H}
 \int^t \left( \frac{H^2}{\dot\phi^2} - \frac{1}{6} \right) \frac{dt}{a^3} ~.
\end{equation}
Therefore
\begin{eqnarray}
\mathbb{N}_{\dot\phi}
 & = & - \frac{a^3 \dot\phi}{H} \int^t \frac{H^2}{\dot\phi^2} \frac{dt}{a^3}
\\
& = & \frac{1}{3 \dot\phi} \left[ 1 + \frac{2 a^3 \dot\phi^2}{H}
 \int^t \left( \frac{\ddot\phi}{H \dot\phi} + \frac{\dot\phi^2}{4H^2} \right)
 \frac{H^2}{\dot\phi^2} \frac{dt}{a^3} \right] ~.
\end{eqnarray}
This gives
\begin{equation}
\ln \mathbb{N}_{\dot\phi}^2
 + \frac{2}{3} \frac{\left(\mathbb{N}_{\dot\phi}\right)'}{\mathbb{N}_{\dot\phi}}
\simeq \ln \left( \frac{1}{3 \dot\phi} \right)^2
- \frac{2}{3} \frac{\ddot\phi}{H \dot\phi} - \frac{1}{3} \frac{\dot\phi^2}{H^2} ~.
\end{equation}
The other contributions to $\Pi$ are evaluated as
\begin{equation}
\ln \left( \frac{3 H^2}{2\pi} \right)^2 + \frac{2}{3} \frac{\left(H^2\right)'}{H^2}
\simeq \ln \left( \frac{3 H^2}{2\pi} \right)^2 + \frac{2}{3} \frac{\dot\phi^2}{H^2}
\end{equation}
and
\begin{equation}
\ln \left( \frac{k}{xaH} \right)^4 \simeq - 2 \frac{\dot\phi^2}{H^2} ~.
\end{equation}
Therefore
\begin{equation}
\ln \Pi^2 + \frac{2}{3} \frac{\Pi'}{\Pi}
\simeq \ln \left( \frac{H^2}{2\pi \dot\phi} \right)^2
 - \frac{2}{3} \frac{\ddot\phi}{H \dot\phi} - \frac{5}{3} \frac{\dot\phi^2}{H^2} ~.
\end{equation}
We see that this is identical to \cite{gsr}
\begin{equation}
\ln \frac{1}{f^2} + \frac{2}{3} \frac{f'}{f}
 \simeq \ln \left( \frac{H^2}{2\pi \dot\phi} \right)^2
 - \frac{2}{3} \frac{\ddot\phi}{H \dot\phi} - \frac{5}{3} \frac{\dot\phi^2}{H^2} ~.
\end{equation}

\section{Configuration space formula}
\label{AppConfig}

For $D$-dimensional field space, the phase space is $2D$-dimensional.
On super-horizon scales during standard slow-roll inflation, however,
half of the solutions are decaying, and eventually the phase space reduces to $D$-dimensions.
Then $\dot\phi$ is uniquely determined at each point in the configuration space,
\mbox{i.e.} $\dot\phi=\dot\phi(\phi)$.
In such a case, we can define $N(\phi) \equiv N(\phi,\dot\phi(\phi))$ with
\begin{equation}
\nabla_\mathbf{a} N
= N_{\phi^\mathbf{a}} + N_{\dot\phi^\mathbf{b}} \nabla_\mathbf{a} \dot\phi^\mathbf{b} ~.
\end{equation}
Substituting into Eq.~(\ref{deltaN}) gives
\begin{equation}
\delta N = \left( \nabla_\mathbf{a} N \right) \delta\phi^\mathbf{a}
+ N_{\dot\phi^\mathbf{a}} \left[ \dot{\delta\phi^\mathbf{a}}
- \dot\phi^\mathbf{a} A
 - \delta\phi^\mathbf{b} \nabla_\mathbf{b} \dot\phi^\mathbf{a} \right]
- \frac{q^2}{9H} X \mathcal{S}_\mathrm{f} ~.
\end{equation}
Note that
\begin{equation}
\Delta^\mathbf{a}\equiv\dot{\delta\phi^\mathbf{a}} - \dot\phi^\mathbf{a} A
 - \delta\phi^\mathbf{b} \nabla_\mathbf{b} \dot\phi^\mathbf{a}
\end{equation}
is gauge invariant, and that it vanishes for the super-horizon adiabatic growing mode of Eq.~(\ref{SHGM}).

More generally, Eq.~(\ref{phieom}) gives
\begin{equation}\label{graddotphi}
3 H \nabla_\mathbf{b} \dot\phi^\mathbf{a}
 = - h^\mathbf{ac} V_{\phi^\mathbf{c}\phi^\mathbf{b}}
 - \frac{1}{2H} \dot\phi^\mathbf{a} V_{\phi^\mathbf{b}}
 - \nabla_\mathbf{b} \ddot\phi^\mathbf{a}
 - \frac{1}{2H} \dot\phi^\mathbf{a} \dot\phi_\mathbf{c}
 \nabla_\mathbf{b} \dot\phi^\mathbf{c} ~,
\end{equation}
and we have
\begin{equation}\label{gradddotphi}
\nabla_\mathbf{b} \ddot\phi^\mathbf{a}
 = - R^\mathbf{a}_\mathbf{\ cdb} \dot\phi^\mathbf{c} \dot\phi^\mathbf{d}
 + \left( \nabla_\mathbf{b} \dot\phi^\mathbf{a} \right)^\centerdot
 + \left( \nabla_\mathbf{b} \dot\phi^\mathbf{c} \right)
  \left( \nabla_\mathbf{c} \dot\phi^\mathbf{a} \right) ~.
\end{equation}
Therefore, using Eqs.~(\ref{A}) and~(\ref{deltaphieom}), we have
\begin{eqnarray}
\dot\Delta^\mathbf{a} + 3 H \Delta^\mathbf{a}
+ \left( \nabla_\mathbf{b} \dot\phi^\mathbf{a}
+ \frac{1}{2H} \dot\phi^\mathbf{a} \dot\phi_\mathbf{b} \right)\Delta^\mathbf{b}
+ q^2 \delta\phi_\mathrm{f}^\mathbf{a} = 0 ~.
\end{eqnarray}

In the case of standard slow-roll, we see from Eqs.~(\ref{graddotphi})
 and~(\ref{gradddotphi}) that
\begin{equation}
\left| \frac{1}{H} \nabla_\mathbf{b} \dot\phi^\mathbf{a} \right| \ll 1 ~.
\end{equation}
Hence $\Delta^\mathbf{a}$ is decaying on super-horizon scales.
Therefore the configuration space formula \cite{SS,Nakamura,JO}
\begin{equation}\label{config}
\delta N = \left( \nabla_\mathbf{a} N \right) \delta\phi^\mathbf{a}
\end{equation}
is valid for standard slow-roll on super-horizon scales.

\end{document}